\documentclass[a4paper,11pt]{article}

\usepackage{hyperref}

\def\be{\begin{equation}}
\def\te{\end{equation}}
\def\ee{\end{equation}}
\def\ba{\begin{eqnarray}}
\def\bea{\begin{eqnarray}}
\def\nn{\nonumber\\}
\def\tea{\end{eqnarray}}
\def\ea{\end{eqnarray}}
\def\eea{\end{eqnarray}}
\begin{document}

\title{Field Theory Approaches to Relativistic Hydrodynamics}

\author{Nahuel Mir\'on-Granese\\
Consejo Nacional de Investigaciones Científicas y Técnicas (CONICET),Argentina,\\ 
Facultad de Ciencias Astronómicas y Geofísicas,\\ Universidad Nacional de La Plata,Argentina,\\ 
and Departamento de Física, Facultad de Ciencias Exactas y Naturales,\\ Universidad de Buenos Aires, Argentina\thanks{nahuelmg@fcaglp.unlp.edu.ar}\\[0.5cm]
Alejandra Kandus\\
LATO-DCET Universidade Estadual de Santa Cruz,Ilhéus-BA, Brazil\thanks{kandus@uesc.br}\\[0.5cm]
Esteban Calzetta\\
Universidad de Buenos Aires, Facultad de Ciencias Exactas y Naturales,\\ Departamento de Física, Argentina,\\ y CONICET - Universidad de Buenos Aires,\\ Instituto de Física de Buenos Aires (IFIBA), Argentina\thanks{calzetta@df.uba.ar}}

\maketitle

\begin{abstract}{Just as non relativistic fluids, oftentimes we find relativistic fluids in situations where random fluctuations cannot be ignored, thermal and turbulent fluctuations being the most relevant examples. Because of the theory's inherent nonlinearity, fluctuations induce deep and complex changes in the dynamics of the system. The Martin-Siggia-Rose technique is a powerful tool that allows us to translate the original hydrodynamic problem into a quantum field theory one, thus taking advantage of the progress in the treatment of quantum fields out of equilibrium. To demonstrate this technique, we shall consider the thermal fluctuations of the spin two modes of a relativistic fluid, in a theory where hydrodynamics is derived by taking moments of the Boltzmann equation under the relaxation time approximation.}
\end{abstract}

\section{Introduction}

The success of hydrodynamics in the description of relativistic heavy ion collisions \cite{RZ13,RR19,EC2013} have brought to the fore the problem of the very early stages of the process, as hydrodynamic behavior appeared on time scales that were not larger than the expected relaxation times \cite{Martinez18,Rom18,MS18,MS18b,Heinz20,Kurkela20,DN21,Silva22}. In a parallel development, the possibility of relativistic viscous fluids playing an important role on the evolution of electromagnetic and gravitational background fields in cosmological and astrophysical settings \cite{BPP,Goswami,Alford,Cao,Hindmarsh,Friedman,NMGEC,NMG} has similarly brought attention to regimes where the relaxation time is not the shortest time scale in the problem.

In this regime, not only the fluid is strongly out of equilibrium, but also its fluctuations are not negligible. To mention two important instances of this phenomenon, these fluctuations may be of thermal origin, or else due to the nonlinear amplification of external influences leading to turbulence \cite{K08,FW11,CR11,F13,AKLN14,ED18,EC1}. 

Field theory methods, through the so-called Martin-Siggia-Rose (MSR) approach \cite{MSR,Kamenev,Eyink,JZEC}, have proven to be a powerful tool in the analysis of fluctuating fluids. The method is based on the construction of a generating functional for the correlation functions of the fluid \cite{Kovtun,KMR,HKR,Haehl15,MT16,Li17,Haehl18,MT20}. This functional has the same form as the generating functional for a non equilibrium quantum field \cite{CH08}, and this brings to bear the substantial tools available for the study of these systems, particularly functional methods based on the one or two particles irreducible effective action. The first choice is relevant when we are interested in the mean values of the hydrodynamic variables, while the second is superior when the goal is the description of correlations, which is the task at hand.

In this paper we shall present the MSR approach to fluctuating relativistic viscous fluids. Hydrodynamics is conceived as an effective theory describing the long lived modes of a more fundamental description, which is usually either a field theory \cite{CH08} (including conformal field theories which may be studied though holographic methods \cite{BRSSS}) or else a kinetic theory \cite{Isr72}. For concreteness we shall assume the latter. We therefore start from a Boltzmann equation (as we shall discuss below, for the kind of problems we have in mind it is enough to work within the relaxation time approximation for the collision term \cite{AW1,AW2,TI10,JPREC,RDN21}) and derive hydrodynamics by taking moments of this equation \cite{DMNR12b,AMR}, in a manner to be fully described below.

As we said above, in applications the emphasis is on the interaction of the fluid with electromagnetic and gravitational fields, and therefore the main interest is on the spin $1$ and $2$ modes of the fluid. We have treated electromagnetic interactions elsewhere (including the possibility of the fluid amplifying a seed field through the Weibel instability \cite{AKEC1,KMGC2}); in this paper we shall focus on spin $2$ modes.

The existence of non hydrodynamic tensor modes in a relativistic fluid is a generic prediction of kinetic theory \cite{HisLind3,Natsuume08,GPEC}. To capture them we must recur to a particular parameterization of the one particle distribution function, including explicitly a second order tensor as an independent hydrodynamic variable; as we shall show below, to provide this mode with a  finite propagation speed we must include a third order tensor as well \cite{GPEC,EC2,BD21}. As a test case for the formalism, we shall formulate a minimal model including a dynamical tensor mode and shall study how nonlinearities modify the spectrum of the thermal fluctuation of this mode. Concretely, we shall show that the equal time thermal spectrum is flat at long wavelengths (as it is in the linear theory) but becomes a power law at short ones, resembling the spectra characteristic of relativistic turbulence \cite{EC1}.

This paper is organized as follows. In the next section we formulate the MSR approach in the language of the two particle irreducible effective action (2PIEA). We conclude by stating the concrete form of the dressed correlations, to be computed in the following.

We then switch to a presentation of stochastic hydrodynamics, as derived from the moments of a stochastic kinetic equation \cite{LL57,LL59,FUI,FUII,ECBLH,StoBol}, after a parameterization of the one particle distribution function which includes, besides temperature and four velocity, the tensor hydrodynamic variables required to capture the spin $2$ non hydrodynamic mode of the fluid \cite{GPEC}. We shall consider a massless relativistic gas, so we shall not include a chemical potential among the hydrodynamic variables.

Finally we deploy the MSR tools to compute the dressed correlations of the tensor modes to one loop accuracy. This requires computing corrections to both the causal propagator (the so-called self energy) and to the noise correlation (the so-called noise kernel). Together they determine the tensor mode correlation. For easier comparison with the correlation of the linear theory, we compute the equal time correlation, thus obtaining the dressed spectrum of spin two fluctuations.

We conclude with some brief final remarks. The details of the calculation of the relevant Feynman graphs are given in the Appendix.

\section{The Martin-Siggia-Rose approach}
\subsection{From Stochastic to Quantum Fields}
Let us begin by reviewing how one can translate a stochastic field theory into a quantum one \cite{MSR,Kamenev,Eyink,JZEC}. One has a theory of fields $X^{\alpha}$ (the $\alpha$ index accounts for space-time, Lorentz and species indexes) obeying nonlinear stochastic equations of motion 

\be
P^{a}=D^{a}_{\beta}X^{\beta}+\Lambda^{a}_{\beta\gamma}X^{\beta}X^{\gamma}+\ldots=F^{a}
\label{stoc}
\te
The $F^{a}$ are assumed to be a Gaussian noise with zero mean and self correlation

\be
\left\langle F^{a}F^{b}\right\rangle=\Phi^{ab}
\te
The fields $X^{\alpha}$ become stochastic themselves, with a probability density functional

\be
\mathcal{X}\left[X^{\alpha}\right]=\int\;DF^{b}\;\mathcal{F}\left[F^{b}\right]\delta\left[X^{\alpha}-X^{\alpha}\left[F^{b}\right]\right]
\te
where $X^{\alpha}\left[F^{b}\right]$ is the  solution to eq. (\ref{stoc}) for a given realization $F^{b}$ of the noise, and

\be
\mathcal{F}\left[F^{b}\right]=C\;\exp\left\{-\frac12F^{b}\Phi^{-1}_{bc}F^{c}\right\}
\te
is the probability density functional for the noise.

For simplicity we assume that the $X^{\alpha}$ fields do not develop a nonzero expectation value. Then the interest lies on the self correlation

\be
G_1^{\alpha\beta}=\left\langle X^{\alpha}X^{\beta}\right\rangle
\te
$G_1^{\alpha\beta}$ may be derived from a generating functional

\be
G_1^{\alpha\beta}=-2i\left.\frac{\delta W\left[K_{\beta\gamma}\right]}{\delta K_{\alpha\beta}}\right|_{K_{\alpha\beta}=0}
\te

\bea
e^{iW\left[K_{\beta\gamma}\right]}&=&\int\;DX^{\alpha}\;\mathcal{X}\left[X^{\alpha}\right]\;\exp\left\{\frac i2X^{\alpha}K_{\alpha\beta}X^{\beta}\right\}\nn
&=&\int\;DX^{\alpha}DF^{b}\;\mathcal{F}\left[F^{b}\right]\delta\left[X^{\alpha}-X^{\alpha}\left[F^{b}\right]\right]\;\exp\left\{\frac i2X^{\alpha}K_{\alpha\beta}X^{\beta}\right\}
\tea
We now proceed as follows. First we use the identity

\be
\delta\left[X^{\alpha}-X^{\alpha}\left[F^{\beta}\right]\right]=Det\left[\frac{\delta P^{a}}{\delta X^{\beta}}\right]\delta\left[P^{a}-F^{a}\right]
\te
The $\delta$-functions may be added to the exponent by introducing auxiliary fields $Y_a$, as well as the determinant by adding ghost fields. It can be shown that ghosts play no role in the discussion below, so we shall simply assume that the determinant is a constant \cite{ZJ}. So we get

\be
e^{iW\left[K_{\beta\gamma}\right]}=\int\;DY_aDX^{\alpha}DF^{b}\;\mathcal{F}\left[F^{b}\right]\;\exp\left\{iY_a\left(P^{a}-F^{a}\right)+\frac i2X^{\alpha}K_{\alpha\beta}X^{\beta}\right\}
\te
Finally we integrate over the noises $F^a$ to get

\be
e^{iW\left[K_{\beta\gamma}\right]}=\int\;DY_aDX^{\alpha}\;\exp\left\{iY_aP^{a}-\frac12Y_a\Phi^{ab}Y_b+\frac i2X^{\alpha}K_{\alpha\beta}X^{\beta}\right\}
\te
At this point it is convenient to think of $\mathbf{X}^A=\left(X^{\alpha},Y_a\right)$ as a single field, and add sources as necessary so that we get a generating functional for the correlations $\mathbf{G}^{AB}=\left\langle \mathbf{X}^a\mathbf{X}^b\right\rangle$, namely

\be
e^{iW\left[K_{\beta\gamma}\right]}=\int\;D\mathbf{X}^A\;\exp\left\{iS\left[\mathbf{X}^A\right]+\frac i2\mathbf{X}^{A}\mathbf{K}_{AB}\mathbf{X}^b\right\}
\te
where

\be
S\left[\mathbf{X}^A\right]=Y_aP^{a}+\frac i2Y_a\Phi^{ab}Y_b
\te
The ``doubling of degrees of freedom'' gives this the structure of the classical action for a quantum field defined on a \emph{closed time-path} \cite{CH08}. By performing a Lagrange transform with respect to $\mathbf{K}_{AB}$ 
we obtain the \emph{two-particle irreducible effective action} (2PIEA)
|
\be
\Gamma\left[\mathbf{G}^{AB}\right]=\left.W\left[\mathbf{K}_{AB}\right]-\frac 12\mathbf{K}_{AB}\mathbf{G}^{AB}\right|_{W_{,\mathbf{K}}=\mathbf{G}/2}
\te
The variation of the 2PIEA yields the Schwinger-Dyson equations

\be
\frac{\delta\Gamma}{\delta\mathbf{G}^{AB}}=0
\te
which are the most efficient way to find the correlations.

{{The effective action approach has points in common with the non equilibrium generating functional introduced by Zubarev \cite{Zub74,Rischke18,Bec19,Torr21}; however, the goal here is not to compute the correlation functions directly from a generating functional, but rather the equations of motion thereof, which are similar to the Schwinger-Dyson equations from field theory}}

{We should emphasize that the reason to appeal to functional methods is to make an efficient use of the information already encoded in the equations of motion. In principle, one could use field theory methods without introducing path integrals, as it was done by Wyld \cite{Wyld61}. However, the path integral formulation makes it easier to implement powerful methods such as the functional renormalization group \cite{Wett93,LC17,Rischke22}, which we aim to discuss in future publications. For further discussion of path integral methods see \cite{Torr22}.}

\subsection{Mining the 2PIEA}

The 2PIEA has the structure \cite{CH08}

\be
\Gamma\left[\mathbf{G}^{AB}\right]=\frac12\left.\frac{\delta^2S}{\delta\mathbf{X}^A\delta\mathbf{X}^B}\right|_{\mathbf{X}^A=0}\mathbf{G}^{AB}-\frac i2\ln\left[Det\; \mathbf{G}^{AB}\right]+\Gamma_Q\left[\mathbf{G}^{AB}\right]
\te
where $\Gamma_Q$ is the sum of all two particle irreducible vacuum graphs with the full propagator $\mathbf{G}^{AB}$ and vertices derived from the interaction action

\be
S_Q=Y_a\left\{\Lambda^{a}_{\beta\gamma}X^{\beta}X^{\gamma}+\ldots\right\}
\te
therefore the Schwinger-Dyson equations

\be
\left.\frac{\delta^2S}{\delta\mathbf{X}^A\delta\mathbf{X}^B}\right|_{\mathbf{X}^A=0}-i\mathbf{G}^{-1}_{AB}+2\frac{\delta\Gamma_Q}{\delta\mathbf{G}^{AB}}=0
\te
The analysis of these equations is greatly simplified by the observation that

\be
\mathbf{G}_{ab}=\left\langle Y_aY_b\right\rangle=0
\te
Then the obvious identity

\be
\left(\begin{array}{cc}\mathbf{G}^{-1}_{\alpha\beta}&\mathbf{G}^{-1b}_{\alpha}\\\mathbf{G}^{-1a}_{\;\;\;\;\;\beta}&\mathbf{G}^{-1ab}\end{array}\right)\left(\begin{array}{cc}\left\langle X^{\beta}X^{\gamma}\right\rangle&\left\langle X^{\beta}Y_c\right\rangle\\\left\langle Y_bX^{\gamma}\right\rangle&0\end{array}\right)=\left(\begin{array}{cc}\delta^{\gamma}_{\alpha}&0\\0&\delta^a_c\end{array}\right)
\te
shows that $\left\langle X^{\beta}Y_c\right\rangle$ is invertible, since

\be
\mathbf{G}^{-1a}_{\;\;\;\;\;\beta}\left\langle X^{\beta}Y_c\right\rangle=\delta^a_c
\label{retarded}
\te
and then the further equation $\mathbf{G}^{-1}_{\alpha\beta}\left\langle X^{\beta}Y_c\right\rangle=0$ shows that $\mathbf{G}^{-1}_{\alpha\beta}=0$. So there are only two families of nontrivial Schwinger-Dyson equations, eq. (\ref{retarded}) and

\be
\mathbf{G}^{-1a}_{\;\;\;\;\;\beta}\left\langle X^{\beta}X^{\gamma}\right\rangle+\mathbf{G}^{-1ab}\left\langle Y_bX^{\gamma}\right\rangle=0
\te
or else, reading the inverse propagators from the Schwinger-Dyson equations

\bea
\left(-i\right)\left[D^a_{\beta}+2\frac{\delta\Gamma_Q}{\delta\left\langle Y_aX^{\beta}\right\rangle}\right]\left\langle X^{\beta}Y_c\right\rangle&=&\delta^a_c\nn
\left(-i\right)\left[D^a_{\beta}+2\frac{\delta\Gamma_Q}{\delta\left\langle Y_aX^{\beta}\right\rangle}\right]\left\langle X^{\beta}X^{\gamma}\right\rangle+\left[\Phi^{ab}-2i\frac{\delta\Gamma_Q}{\delta\left\langle Y_aY_b\right\rangle}\right]\left\langle Y_bX^{\gamma}\right\rangle&=&0
\tea
The propagator

\be
\left\langle X^{\beta}Y_c\right\rangle=i\left\langle \frac{\delta X^{\beta}}{\delta F^c}\right\rangle
\te
is causal. It is the retarded propagator of the theory. By symmetry, $\left\langle Y_bX^{\gamma}\right\rangle$ is the advanced propagator. The propagator $\left\langle X^{\beta}X^{\gamma}\right\rangle=G_1^{\beta\gamma}$ is the physical correlation function of the theory. We now see that we could derive $G_1^{\beta\gamma}$ from a stochastic equation

\be
\left[D^a_{\beta}+\Sigma^a_{\beta}\right]X^{\beta}=F_{dressed}^a
\te
where the self-energy

\be
\Sigma^a_{\beta}=2\frac{\delta\Gamma_Q}{\delta\left\langle Y_aX^{\beta}\right\rangle}
\te
and the dressed noise has a self correlation

\be
\left\langle F_{dressed}^aF_{dressed}^b\right\rangle\equiv N^{ab}=\Phi^{ab}-2i\frac{\delta\Gamma_Q}{\delta\left\langle Y_aY_b\right\rangle}
\te
$N^{ab}$ is the so-called \emph{noise kernel}.

\subsection{The lowest order correlation}

To make the analysis above more concrete, we shall compute the lowest order correction to the self-energy and to the noise kernel. Keeping only the quadratic terms in the original stochastic equation, and assuming $\Lambda^a_{\beta\gamma}$ is symmetric on $\left(\beta,\gamma\right)$, the lowest order contribution to $\Gamma_Q$ is

\bea
\Gamma_Q&=&\frac{i}2\Lambda^a_{\beta\gamma}\Lambda^{a'}_{\beta'\gamma'}\left\langle\left( Y_aX^{\beta}X^{\gamma}\right)\left( Y_{a'}X^{\beta'}X^{\gamma'}\right)\right\rangle_{2PI}\nn
&=&2i\Lambda^a_{\beta\gamma}\Lambda^{a'}_{\beta'\gamma'}\left\langle Y_aX^{\beta'}\right\rangle\left\langle X^{\beta}Y_{a'}\right\rangle\left\langle X^{\gamma}X^{\gamma'}\right\rangle\nn
&+&i\Lambda^a_{\beta\gamma}\Lambda^{a'}_{\beta'\gamma'}\left\langle Y_aY_{a'}\right\rangle\left\langle X^{\beta}X^{\beta'}\right\rangle\left\langle X^{\gamma}X^{\gamma'}\right\rangle
\label{lowest}
\tea
and so

\bea
\Sigma^a_{\beta}&=&4i\Lambda^a_{\beta'\gamma}\Lambda^{a'}_{\beta\gamma'}\left\langle X^{\beta'}Y_{a'}\right\rangle\left\langle X^{\gamma}X^{\gamma'}\right\rangle\nn
N^{ab}&=&\Phi^{ab}+2\Lambda^a_{\beta\gamma}\Lambda^{b}_{\beta'\gamma'}\left\langle X^{\beta}X^{\beta'}\right\rangle\left\langle X^{\gamma}X^{\gamma'}\right\rangle
\label{sigmaN}
\tea
In the right hand side of (\ref{sigmaN}) we may use the lowest order propagators 

\bea
\left\langle X^{\beta'}Y_{a'}\right\rangle&=&i\left[D^{-1}\right]^{\beta'}_{a'}\nn
\left\langle X^{\beta}X^{\beta'}\right\rangle&=&\left(-1\right)\left[D^{-1}\right]^{\beta}_{a}\,\Phi^{aa'}\,\left[D^{-1}\right]^{\beta'}_{a'}
\tea
Finally the dressed propagators read
\bea
\left\langle X^{\beta'}Y_{a'}\right\rangle_{dressed}=i\left[D+\Sigma\right]^{-1}{}^{\beta'}_{a'}
\tea
\bea
\left\langle X^{\beta}X^{\beta'}\right\rangle_{dressed}=\left(-1\right)\left[D+\Sigma\right]^{-1}{}^{\beta}_{a}\;N^{aa'}\;\left[D+\Sigma\right]^{-1}{}^{\beta'}_{a'}\\
=\left\langle X^{\beta}Y_{a}\right\rangle_{dressed}\;N^{aa'}\;\left\langle Y_{a'}X^{\beta'}\right\rangle_{dressed}
\tea

\section{From Stochastic kinetic theory to Stochastic hydrodynamics}

\subsection{Relativistic kinetic theory} 

Hydrodynamics is usually conceived as an effective theory which captures the dynamics of the long lived modes of a more fundamental description \cite{Roma666,KurWie}, in practice either field theory (including holographic models) or kinetic theory. In this article we shall take the latter viewpoint, and so it is convenient to start with a brief comment on relativistic kinetic theory \cite{Isr72}.

We shall consider the kinetic theory of massless, neutral particles. They are described by the one-particle distribution function $f\left(x^{\mu},p_{\nu}\right)$, where $p^2=0$ and $p^0\ge 0$. The energy-momentum tensor is

\be
T^{\mu\nu}=\int\;Dp\;p^{\mu}p^{\nu}f
\te
and the entropy

\be
S^{\mu}=\int\;Dp\;p^{\mu}f\left[1-\ln f\right]
\te
Here

\be
Dp=\frac{2d^4p_{\nu}}{\left(2\pi\right)^3}\delta\left(-p^2\right)\theta\left(p^0\right)
\te
is the invariant momentum space measure. Once $T^{\mu\nu}$ is given, we define the fluid velocity $u^{\mu}$, temperature $T$ and inverse temperature vector $\beta^{\mu}=u^{\mu}/T$ from the Landau-Lifshitz prescription $T^{\mu\nu}u_{\nu}=-\rho u^{\mu}$, where $\rho=3T^4/\pi^2$ is the energy density and $u^2=-1$.  The equation of motion is Boltzmann's

\be
p^{\mu}\frac{\partial f}{\partial X^{\mu}}=I_{col}.
\label{Boltzmann}
\te
The collision integral is restricted by energy-momentum conservation

\be
\int\;Dp\;p^{\mu}\;I_{col}=0
\label{momcon}
\te
and the $H$ theorem

\be
\int\;Dp\;\ln f\;I_{col}\le 0
\label{Htheor}
\te
for \emph{any} solution of eq. (\ref{Boltzmann}); this enforces the Second Law $S^{\mu}_{,\mu}=\sigma\ge 0$.

We shall assume that the collision integral expanded to linear order around an equilibrium solution defines a symmetric operator on the space of linear perturbations to the one particle distribution function. Then, because of (\ref{momcon}), this operator must have four null eigenvectors associated to the momenta $p^{\mu}$. Since we are considering a massless gas, we do not enforce particle number conservation. We assume these are the only null eigenvectors, they are the hydrodynamic modes.

The rest of the eigenvectors to the collision operator have negative eigenvalues. We shall consider ``hard'' collision terms, where there is a finite gap between zero and the first nonzero eigenvalue, as opposed to ``soft'' collision terms where there is a continuous spectrum stretching away from zero \cite{Dudynski,LuoYu,DN22}. For the present discussion, a hard collision term may be accurately approximated by an Anderson-Witting or relaxation time approximation collision term \cite{AW1,AW2,AJ14,AJ15,CC15,SB21}

\be
I_{col}=\frac{-1}{\tau}\left(-u_{\sigma}p^{\sigma}\right)\left[f-f_0\right]
\label{AW}
\te
where $f_0$ is an equilibrium solution. Momentum conservation requires $f_0$ to be the equilibrium distribution built from the inverse temperature vector derived from $f$, and $u^{\mu}$ to be corresponding  Landau-Lifshitz velocity.

{Of course this is not the only possible linear approximation to the collision term, but just one of the best known, together with Marle's \cite{Marle1}. Over time other proposals have been advanced, with the goal of allowing for a momentum dependence of the relaxation time \cite{JPREC,KurWie,RDN21,Dash22}, and/or to account for both elastic and inelastic collisions \cite{FR16}.}

The non null eigenvectors of the collision operator are associated to the non hydrodynamic modes. The existence of a spin 2 non hydrodynamic mode is a generic prediction of kinetic theory \cite{GPEC}. For example, if the velocity lies in the $z$ direction, then a perturbation proportional to $p_xp_y$ would contribute to the spin 2 part of the energy momentum tensor. This perturbation must have some nontrivial expansion in terms of collision operator eigenvectors, since it is orthogonal to the hydrodynamic modes.

Most importantly, tensor modes mediate the interaction between the fluid and gravitational waves, both in cosmological scenarios such as the post-inflationary Universe \cite{Goswami,NMGEC,NMG} or the phase transitions era \cite{Hindmarsh} and in astrophysical scenarios such as rotating compact objects \cite{Friedman,Alford} and merging neutron stars \cite{BPP} . Therefore we must understand the dynamics of those modes to correctly describe these phenomena.

\subsection{The moments approach to hydrodynamics}

To obtain hydrodynamics from kinetic theory, we begin by expanding $\ln f$ in a set of functions $f_{\alpha}\left(x^{\mu},p_{\nu}\right)$ \cite{LCEC}

\be
\ln f=\sum_{\alpha}X^{\alpha}f_{\alpha}\left(x^{\mu},p_{\nu}\right)
\te
It is customary to choose $\beta^{\mu}$ as one of the $X^{\alpha}$, with $p^{\mu}$ as the corresponding $f_{\alpha}$. Hydrodynamics follows from the truncation of this development to a (hopefully) few terms; the $X^{\alpha}$ then become the hydrodynamic variables. To obtain the hydrodynamic equations we take moments of the Boltzmann equation with suitable functions $R^a$

\be
\int\;Dp\;R^a\left(x^{\nu},p_{\nu}\right)\left[p^{\mu}\frac{\partial f}{\partial x^{\mu}}-I_{col}\right]=0
\te
The problem is that the truncated $f$ is not a solution of the Boltzmann equation and so we cannot appeal to the $H$ theorem to enforce the Second Law. The way out is to choose the $R^a$ as the $f_{\alpha}$ themselves. In particular, energy-momentum conservation $T^{\mu\nu}_{,\nu}=0$ becomes one the equations of motion.

It is clear that this scheme is still too general; to proceed, we must consider a particular realization. In this work, we shall restrict ourselves to

\be
\ln f=\beta_{\mu}p^{\nu}+X_{\mu\nu}\frac{p^{\mu}p^{\nu}}{\left(-u_{\sigma}p^{\sigma}\right)}+X_{\mu\nu\rho}\frac{p^{\mu}p^{\nu}p^{\rho}}{\left(-u_{\sigma}p^{\sigma}\right)^2}
\label{ansatz}
\te
It is assumed that the $X_{\mu\nu}$ and $X_{\mu\nu\rho}$ tensors are totally symmetric, transverse with respect to $u^{\mu}$ and traceless on any two indexes. 
The $X^{\mu\nu}$ field captures the tensor mode which is our main concern; the $X^{\mu\nu\rho}$ field is then necessary to obtain a nontrivial dynamics for those modes \cite{GPEC,BD21}. The model where $X^{\mu\nu\rho}$ is not included has been analyzed in \cite{KMGC1}. 

The powers of $-u_{\sigma}p^{\sigma}$ in the denominators are included to avoid the non equilibrium terms dominating the equilibrium one. If these powers are not included, the theory becomes a divergence-type model \cite{LMR86,GL90,GL91,ReNa95,ReNa97,LeReRu18}, but the momentum integrals become divergent and must be renormalized \cite{MAEC}. In any case, the application of these models to Bjorken and Gubser flows, where exact solutions to kinetic theory are available, and to relativistic shock waves, shows that including the denominators in (\ref{ansatz}) greatly improves the concordance with the exact results \cite{LCEC,EC2}.

The corresponding equations of motion are

\bea
S^{\mu\nu}_{\mu'\nu'}\int\;Dp\;\frac{p^{\mu'}p^{\nu'}}{\left(-u_{\sigma}p^{\sigma}\right)}\left[p^{\lambda}\frac{\partial f}{\partial x^{\lambda}}-I_{col}\right]&=&0\nn
S^{\mu\nu\rho}_{\mu'\nu'\rho'}\int\;Dp\;\frac{p^{\mu'}p^{\nu'}p^{\rho'}}{\left(-u_{\sigma}p^{\sigma}\right)^2}\left[p^{\lambda}\frac{\partial f}{\partial x^{\lambda}}-I_{col}\right]&=&0
\tea
$S^{\mu\nu}_{\mu'\nu'}$ and $S^{\mu\nu\rho}_{\mu'\nu'\rho'}$ are projectors over the space of totally symmetric, transverse with respect to $u^{\mu}$ and traceless tensors. They can be built from the projector $\Delta^{\mu\nu}=\eta^{\mu\nu}+u^{\mu}u^{\nu}$.

With an integration by parts we may write

\bea
S^{\mu\nu}_{\mu'\nu'}\left[A^{\mu'\nu'\lambda}_{,\lambda}-K^{\mu'\nu'\sigma\lambda}u_{\sigma,\lambda}-I^{\mu'\nu'}\right]&=&0\nn
S^{\mu\nu\rho}_{\mu'\nu'\rho'}\left[A^{\mu'\nu'\rho'\lambda}_{,\lambda}-2K^{\mu'\nu'\rho'\sigma\lambda}u_{\sigma,\lambda}-I^{\mu'\nu'\rho'}\right]&=&0
\tea
where

\bea
A^{\mu'\nu'\lambda}&=&\int\;Dp\;\frac{p^{\mu'}p^{\nu'}}{\left(-u_{\sigma}p^{\sigma}\right)}p^{\lambda} f\nn
K^{\mu'\nu'\sigma\lambda}&=&A^{\mu'\nu'\sigma\lambda}=\int\;Dp\;\frac{p^{\mu'}p^{\nu'}}{\left(-u_{\sigma'}p^{\sigma'}\right)^2}p^{\lambda}p^{\sigma} f\nn
I^{\mu'\nu'}&=&\int\;Dp\;\frac{p^{\mu'}p^{\nu'}}{\left(-u_{\sigma}p^{\sigma}\right)}I_{col}\nn
K^{\mu'\nu'\rho'\sigma\lambda}&=&\int\;Dp\;\frac{p^{\mu'}p^{\nu'}p^{\rho'}}{\left(-u_{\sigma'}p^{\sigma'}\right)^3}p^{\lambda}p^{\sigma} f\nn
I^{\mu'\nu'\rho'}&=&\int\;Dp\;\frac{p^{\mu'}p^{\nu'}p^{\rho'}}{\left(-u_{\sigma}p^{\sigma}\right)^2}I_{col}
\tea
With the Anderson-Witting collision term (\ref{AW}) we get

\bea
I^{\mu'\nu'}&=&\frac{-1}{\tau}\left[T^{\mu'\nu'}-T_0^{\mu'\nu'}\right]\nn
I^{\mu'\nu'\rho'}&=&\frac{-1}{\tau}\left[A^{\mu'\nu'\rho'}-A_0^{\mu'\nu'\rho'}\right]\nn
\tea
The second terms, where $f$ is replaced by $f_0$, will not survive the projector operators and may be discarded.

\subsection{Stochastic hydrodynamics}

The above framework is incomplete in that it does not account for thermal fluctuations. We may fix that by adding a noise term to the Boltzmann equation as  derived from fluctuation-dissipation considerations \cite{LL57,LL59,FUI,FUII,ECBLH,StoBol}.

Let us first consider the fluctuation-dissipation theorem in an abstract setting. Suppose a theory with variables $x^{\alpha}$ whose probability density, in equilibrium, takes the form

\be 
f\left(x^{\alpha}\right)=e^{\Phi\left(x^{\alpha}\right)}
\label{thermaleq}
\te 
For example, if the system is in thermal equilibrium then the potential $\Phi$ is  $-F/T$, where $F$ is the free energy $F=E-TS$. 

Eq. (\ref{thermaleq}) implies that the system is not just sitting at the equilibrium state, which is the maximum of the potential which we identify as $x^{\alpha}=0$, but fluctuates around it, in a way which is prescribed by the classical equipartition theorem

\be
\langle x^{\alpha}J_{\beta}\rangle=\delta^{\alpha}_{\beta}
\te 
where 

\be 
J_{\beta}=-\frac{\partial\Phi}{\partial x^{\beta}}
\te 
is the so-called thermodynamic force. 

The dynamics of the system has a deterministic and a random component

\be
\dot x^{\alpha}=F^{\alpha}_{det}+\zeta^{\alpha}
\label{Langevin}
\te

The deterministic  part drags the system towards $x^{\alpha}=0$; for linear deviations from equilibrium it may be parameterized as

\be
F^{\alpha}_{det}=-\gamma^{\alpha\beta}J_{\beta}
\te 
where the matrix $\gamma^{\alpha\beta}$ is positive definite. Then the fluctuation-dissipation theorem states that

\be 
\langle \zeta^{\alpha}\zeta^{\beta}\rangle=\gamma^{\alpha\beta}+\gamma^{\beta\alpha}
\label{FDT}
\te
Let us apply this general scheme to kinetic theory.
The relevant thermodynamic potential is the Massieu function

\be
\Phi =-\int\;d^3x\;U_{\mu}\Phi^{\mu}
\label{GMF}
\te
where $U^{\mu}$ is the equilibrium four velocity, namely the velocity of an observer at rest with respect to the thermal bath (which we take as $\left(1,0,0,0\right)$) \cite{Gav22b}
\be
\Phi^{\mu}=S^{\mu}+\beta_{0\nu}T^{\mu\nu}=-\int\;Dp\;p^{\mu}f\left[\ln f-1-\beta_{0\nu}p^{\nu}\right]
\te
Upon a perturbation $f=f_{0}+\delta f$ we have {\cite{Gav22}}

\bea
\Phi^{\mu}&=&-\int\;Dp\;p^{\mu}\left(f_0+\delta_f\right)\left[\frac{\delta f}{f_0}-\frac12\left(\frac{\delta f}{f_0}\right)^2-1\right]\nn
&=&\Phi_0^{\mu}-\frac12\int\;Dp\;p^{\mu}\frac{\delta f^2}{f_0}
\tea
The thermodynamic force is derived from the variational derivative of the global Massieu function (\ref{GMF}). 

\be
J\left(\left(t,\vec x\right),\vec p\right)=-\frac{\delta\Phi}{\delta f\left(\left(t,\vec x\right),\vec p\right)}=\frac{1}{\left(2\pi\right)^3}\frac{\delta f}{f_0}\left(\left(t,\vec x\right),\vec p\right)
\te

To write the Anderson-Witting collision term (\ref{AW}) to the required order we need to compute the corrections to $f_{0}$. We start from

\be
T^{\mu\nu}=T_{0}^{\mu\nu}+\delta T^{\mu\nu}
\te 
so, writing $u^{\mu}=U^{\mu}+\delta u^{\mu}$, $T=T_0+\delta T$, $U_{\mu}\delta u^{\mu}=0$ we have

\be
\left[T_{0}^{\mu\nu}+\delta T^{\mu\nu}\right]\left[U_{\nu}+\delta u_{\nu}\right]=-\left[\rho_0 +\delta\rho\right]\left[U^{\mu}+\delta u^{\mu}\right]
\te 
To first order

\be
T_{0}^{\mu\nu}\delta u_{\nu}+\delta T^{\mu\nu}U_{\nu}=-\rho_0\delta u^{\mu}-\delta\rho U^{\mu}
\te
Therefore

\bea
\delta {\rho}&=&\delta T^{\mu\nu}U_{\mu}U_{\nu}\nn
\delta u^{\mu}&=&-\frac {\Delta^{\mu}_{\rho}\delta T^{\rho\nu}U_{\nu}}{\rho_0+p_0}
\tea
where we have used that $T_0^{\mu\nu}=\rho_0U^{\mu}U^{\nu}+p_0\Delta^{\mu\nu}$, $\Delta^{\mu\nu}=\eta^{\mu\nu}+U^{\mu}U^{\nu}$ and finally, using the Stefan-Boltzmann relation,

\be
\frac{\delta T}{T_0}=\frac14\frac{\delta\rho}{\rho_0}
\te

\be
\delta\beta_{\nu}=\frac1{T_0}\delta u_{\nu}-\frac{U_{\nu}}{T_0^2}\delta T=-\frac {\Delta^{\nu}_{\rho}\delta T^{\rho\sigma}U_{\sigma}}{T_0\left(\rho_0+p_0\right)}-\frac{U_{\nu}}{4T_0\rho_0}\delta T^{\rho\sigma}U_{\rho}U_{\sigma}
\te
Putting all together

\bea
\delta I_{col}&=&\frac1{\tau}U_{\mu}p^{\mu}\left[\delta f-\delta f_{0}\right]\nn
&=&-\left(2\pi\right)^3\frac{p^0}{\tau}f_0\left[J+\frac3{4\rho_0T_0}p^{\nu}\left[\Delta_{\nu\rho}+\frac1{3}U_{\nu}U_{\rho}\right]\int\;Dp'\;p'^{\rho}U_{\sigma}p'^{\sigma}f_0J\right]
\tea
where we have used the equation of state $p_0 = \frac{\rho_0}{3}$. 

In summary, if we assume the collision integral acquires an stochastic component

\be
I_{col}\to I_{col}+\mathcal{I}\left(\left(t,x^j\right),p_j\right)
\te 
then from the fluctuation-dissipation theorem (\ref{FDT})

\bea
&&\left\langle \mathcal{I}\left(\left(t,x^j\right),p_j\right)\mathcal{I}\left(\left(t',y^k\right),q_k\right)\right\rangle=\frac2{\tau}\left(-u_{\mu}p^{\mu}\right)\left(-u_{\mu}q^{\mu}\right)f_0\left(p\right)\delta\left(x-y\right)\delta\left(t-t'\right)\nn
&&\left\{\left(2\pi\right)^3\delta\left(p-q\right)-\frac{3f_0\left(q\right)p^{\nu}q^{\rho}}{4\rho_0T_0}\left[\Delta_{\nu\rho}+\frac1{3}u_{\nu}u_{\rho}\right]\right\}.
\tea
After taking moments, the hydrodynamic equations get noise terms \cite{EC98,KMGC1}
 
\bea
\mathcal{I}^{\mu}&=&\int\;Dp\;{p^{\mu}}\mathcal{I}\nn
\mathcal{I}^{\mu\nu}&=&S^{\mu\nu}_{\mu'\nu'}\int\;Dp\;\frac{p^{\mu'}p^{\nu'}}{\left(-u_{\sigma}p^{\sigma}\right)}\mathcal{I}\nn
\mathcal{I}^{\mu\nu\rho}&=&S^{\mu\nu\rho}_{\mu'\nu'\rho'}\int\;Dp\;\frac{p^{\mu'}p^{\nu'}p^{\rho'}}{\left(-u_{\sigma}p^{\sigma}\right)^2}\mathcal{I}
\tea
Now we find

\be
\left\langle \mathcal{I}^{\mu}\left(x\right)\mathcal{I}^{\nu}\left(x'\right)\right\rangle=\left\langle \mathcal{I}^{\mu}\left(x\right)\mathcal{I}^{\nu\rho}\left(x'\right)\right\rangle=\left\langle \mathcal{I}^{\mu}\left(x\right)\mathcal{I}^{\nu\rho\sigma}\left(x'\right)\right\rangle=0
\te
so we may simply take $\mathcal{I}^{\mu}=0$. The noise does not feed energy, but entropy, into the system \cite{ED18,EC1}.

The remaining correlations are

\bea
\left\langle \mathcal{I}^{\mu\nu}\left(x\right)\mathcal{I}_{\lambda\tau}\left(x'\right)\right\rangle&=&\frac4{15\tau}\tilde\rho S^{\mu\nu}_{\lambda\tau}\delta\left(x-x'\right)\nn
\left\langle \mathcal{I}^{\mu\nu}\left(x\right)\mathcal{I}_{\lambda\tau\omega}\left(x'\right)\right\rangle&=&0\nn
\left\langle \mathcal{I}^{\mu\nu\rho}\left(x\right)\mathcal{I}_{\lambda\tau\omega}\left(x'\right)\right\rangle&=&\frac4{35\tau}\tilde\rho S^{\mu\nu\rho}_{\lambda\tau\omega}\delta\left(x-x'\right)
\tea
where $\tilde\rho=12T^5/\pi^2$.

{We observe that the noise in the hydrodynamic equations is additive, but for a more realistic collision integral \cite{MS18c,Mullins22} there will be multiplicative noise too \cite{Arnold00,DB10}. We aim to discuss this issue in future publications.}

\subsection{MSR Hydrodynamics}

To set up the MSR action corresponding to a viscous relativistic fluids, we introduce Lagrange multipliers $Y_{\mu}$, $Y_{\mu\nu}$ and $Y_{\mu\nu\rho}$, the latter being transverse and traceless. Then the action reads

\bea
S&=&\int\;d^4x\;\left\{-Y_{\mu,\nu}T^{\mu\nu}-Y_{\mu\nu,\rho}A^{\mu\nu\rho}-Y_{\mu\nu}\left[K^{\mu\nu\sigma\lambda}u_{\sigma,\lambda}+I^{\mu\nu}\right]-Y_{\mu\nu\rho,\sigma}A^{\mu\nu\rho\sigma}\right.\nn
&-&\left.Y_{\mu\nu\rho}\left[2K^{\mu\nu\rho\sigma\lambda}u_{\sigma,\lambda}+I^{\mu\nu\rho}\right]
+2i\frac{\tilde\rho}{\tau}\left[\frac1{15}Y^{\mu\nu}Y_{\mu\nu}+\frac1{35}Y^{\mu\nu\rho}Y_{\mu\nu\rho}\right]\right\}
\tea
So far the treatment is fully nonlinear. However, we know from the Navier-Stokes equations that the most relevant nonlinear terms are those related to convective derivatives, over and above corrections to the viscous energy momentum tensor. To capture that kind of behavior we shall linearize on $X^{\mu\nu}$ and $X^{\mu\nu\rho}$, while leaving $u^{\mu}$ arbitrary, and then we define

\be
u^{\mu}=\mu U^{\mu}+v^{\mu}
\te
where $U^{\mu}=\delta^{\mu}_0$, $U^{\mu}v_{\mu}=0$, $\mu=1+\frac12v^{\mu}v_{\mu}+$ higher order. Then $\Delta^{00}=v^kv_k$, $\Delta^{0k}=\mu v^k$, $\Delta^{jk}=\delta^{jk}+v^jv^k$. Observe that similarly $U_{\mu}X^{\mu\nu}=-v_{\mu}X^{\mu\nu}$ is a second order quantity, while $U_{\mu}U_{\nu}X^{\mu\nu}=v_{\mu}v_{\nu}X^{\mu\nu}$ is of third order. 

Given the complexity of the theory, we shall produce a demonstrative calculation retaining only some of the relevant Feynman graphs.

Our goal is to see how radiative corrections affect the tensor fluctuations. We cannot build the theory out of tensor modes alone, because tensor modes couple to each other through vector modes. The simplest non trivial theory has two vector modes and two tensor modes, namely, the vector part of $v^k$ (therefore, we assume that $v^k_{,k}=0$), the vector and tensor parts of $X^{jk}$ (to single out which, we assume $X_{jk}=x_{j,k}+x_{k,j}+\bar x_{jk}$, with $x^j_{,j}=\bar x^{jk}_{,k}=0$) and the tensor part of
$X_{jkl}={\bar x}'_{jk,l}+{\bar x}'_{kl,j}+{\bar x}'_{lj,k}$ where ${\bar x'}\;^{jk}_{,k}=0$. To obtain equations for them we perform a similar decomposition of the Lagrange multipliers, namely $Y_j=y_j$, $Y_{jk}=y'_{j,k}+y'_{k,j}+\bar y_{jk}$, and $Y_{jkl}={\bar y}'_{jk,l}+{\bar y}'_{kl,j}+{\bar y}'_{lj,k}$. Because of rotation invariance, there are no nontrivial correlations between vector and tensor variables. Finally, we shall keep only one interaction, namely the coupling between $v^k$, $\bar x^{jk}$ and $\bar y^{jk}$ which comes from the convective term. Now the action reads

\bea
S&=&\int\;d^4x\;\left\{y_{j}\left[\frac43\rho \dot v^j+\frac2{15}\tilde\rho\;\mathbf{\Delta} x^j\right]\right.\nn
&-&2y'_{j}\mathbf{\Delta}\left[\frac2{15}\tilde\rho \left[\dot x^{j}+\frac1{\tau}x^{j}\right]+\frac{4\rho}{15}v^{j}\right]\nn
&+&\bar y_{jk}\left[\frac2{15}\tilde\rho \left[\dot {\bar x}^{jk}+\frac1{\tau}\bar x^{jk}\right]+\frac2{35}\tilde\rho\;\mathbf{\Delta} {\bar x}^{'jk}\right]\nn
&-&\bar y'_{jk}\frac6{35}\tilde\rho  \mathbf{\Delta}\left[\dot {\bar x}^{'jk}+\frac1{\tau}\bar x^{'jk}+\frac13\bar x^{jk}\right]\nn
&+&\left.2i\frac{\tilde\rho}{15\tau}\left[-2y'_j\mathbf{\Delta}y'_j+\bar y_{jk}\bar y_{jk}\right]-2i\frac{\tilde\rho}{35\tau}y^{'jk}\mathbf{\Delta} y'_{jk}+\bar y_{jk}\frac2{15}\tilde\rho v^{l} \bar x^{jk}_{,l}\right\}
\tea
Integrating out the $x^j$ and $\bar x^{'jk}$ fields we get the constraints

\bea
\frac2{15}\tilde\rho\;\mathbf{\Delta} y_{j}+\frac4{15}\tilde\rho \mathbf{\Delta}\left[\dot y'_{j}-\frac1{\tau}y'_{j}\right]&=&0\nn
\frac2{35}\tilde\rho\;\mathbf{\Delta} \bar y_{jk}+\frac6{35}\tilde\rho  \mathbf{\Delta}\left[\dot y'_{jk}-\frac1{\tau}y'_{jk}\right]&=&0
\tea
We use these constraints to elliminate $y_j$ and $\bar y_{jk}$, whereby

\bea
S&=&\int\;d^4x\;\left\{\frac83\rho y'_{j} \left[\left(\frac {\partial}{\partial t} +\frac1{\tau}\right)\dot v^j-\frac15\mathbf{\Delta}v^{j}\right]\right.\nn
&+&\frac2{5}\tilde\rho y'_{jk}\left[\left(\frac {\partial}{\partial t} +\frac1{\tau}\right)^2\bar x^{jk}-\frac17\mathbf{\Delta}\bar x^{jk}\right]\nn
&+&2i\frac{\tilde\rho}{15\tau}\left[-2y'_j\mathbf{\Delta}y'_j+9\left[\frac{\partial}{\partial t} -\frac1{\tau}\right]y'_{jk}\left[\frac{\partial}{\partial t} -\frac1{\tau}\right]y^{'jk}-\frac97y^{'jk}\mathbf{\Delta} y'_{jk}\right]\nn
&-&\left.\frac2{5}\tilde\rho \left[\dot y'_{jk}-\frac1{\tau}y'_{jk}\right]v^{l} \bar x^{jk}_{,l}\right\}
\label{action}
\tea

\section{Perturbative evaluation of the 2PIEA}

We shall now evaluate the corrections to the ``classical'' correlations derived from the action (\ref{action}), to first order in the loop expansion. 

We are building graphs with three kinds of internal lines, corresponding to the velocity symmetric correlation $\left\langle vv\right\rangle$, the tensor symmetric correlation $\left\langle \bar x\bar x\right\rangle$ and the tensor causal propagator $\left\langle \bar xy'\right\rangle$, cubic vertices where the incoming lines are one of each kind, and external lines which may be of two types, $\bar x$ or $y'$. We shall be interested in contributions to the self-energy, whereby one external line is of $y'_{jk}$ type and the other of $\bar x^{jk}$, and corrections to the noise kernel, where both external lines are of the $y'_{jk}$ kind. Therefore we have the relationships

\bea
V&=&2I_{VV}\nn
V&=&I_{XY}+2I_{XX}+E_X\nn
V&=&I_{XY}+E_Y\nn
I_{VV}+I_{XX}+I_{XY}-V&=&L-1
\tea
where $V$ is the number of vertices, $L$ of loops, $I_{XX}$ the internal lines of the $XX$ kind, and $E_X$ the external lines of the $X$ kind. The solution to this system reads

\bea
I_{VV}&=&L-1+\frac12E_X+\frac12E_Y\nn
I_{XX}&=&-\frac12E_X+\frac12E_Y\nn
I_{XY}&=&2\left(L-1\right)+E_X\nn
V&=&2\left(L-1\right)+E_X+E_Y
\tea
 As $\tau\to\infty$, the poles of the propagators move closer to the real axis, and this causes each loop integral to diverge linearly in $\tau$ in the free streaming limit.  Therefore we estimate the contribution of each loop integral as $\left(k\tau\right)k^4$. Including this factor, a given graph scales as

\be
\left(\frac{\tilde\rho}{\tau\rho^2 k^2}\right)^{I_{VV}}\left(\frac1{\tau\tilde\rho k^2}\right)^{I_{XX}}\left(\frac1{\tilde\rho k^2}\right)^{I_{XY}}\left(\tilde\rho k^2\right)^V\left(\tau k^{5}\right)^{L}
\te
Rearranging we get

\be
\left(\frac{\tau\rho^2 k^2}{\tilde\rho}\right)\left(\frac{\tilde\rho }{\rho }\right)^{E_X}\left(\frac{\tilde\rho}{\tau\rho}\right)^{E_Y}\left(\frac{\tilde\rho k^3}{\rho^2}\right)^L
\te
Self energy graphs have $E_X=E_Y=1$, so

\be
\Sigma\approx \tilde\rho k^2\left(\frac{\tilde\rho k^3}{\rho^2}\right)^L
\label{sigmascaling}
\te
A noise kernel graph has $E_X=0$, $E_Y=2$

\be
N\approx\left(\frac{\tilde\rho k^2}{\tau}\right)\left(\frac{\tilde\rho k^3}{\rho^2}\right)^L
\label{noisescaling}
\te
 We see that the loop expansion is reliable for all momenta $k\le T$. In the following we shall consider the first corrections to the self energy and the noise kernel as we approach the free streaming limit $\tau\to\infty$.

 {In the opposite limit $\tau\to 0$ the tensor modes disappear and the equations for the scalar and vector modes become constitutive relations appropriate to the Chapman-Enskog theory. Therefore in that limit we recover the analysis of refs. \cite{KMR11,JPREC12}. }

\subsection{``Classical'' propagators}

We define the Fourier transform in both time and space as
\bea
h^j(x)=\int \frac{d\omega}{2\pi}\,\frac{d^3k}{(2\pi)^3}e^{-i\left(\omega\, t-\vec k\cdot \vec x\right)} \,h^j(\omega,\vec k)
\tea
Due to isotropy and time-translation symmetry, the vector propagators read (where $h^i$ and $g^j$ are just two generic divergenceless vector fields)
\bea
\left\langle h^i(x)\,g^j(x')\right\rangle=\left\langle h^ig^j\right\rangle(t-t',\vec x-\vec x')
\eea
and consequently,
\bea
\left\langle h^i(\omega,\vec k)\,g^j(\omega',\vec k')\right\rangle=(2\pi)^4\, \delta(\vec k+\vec k')\,\delta(\omega+\omega')\,\langle h^ig^j\rangle(\omega,\vec k)
\eea
where
\bea
\langle h^ig^j\rangle(\omega,\vec k)= G_{hg}(\omega,k)\,P^{ij}(\hat k)
\tea
with
\bea
P^{ij}(\hat k)=\delta^{ij}-\frac{k^ik^j}{k^2}\,
\eea
the vector spatial projector. In case of having tensor propagators $P^{ij}$ must be replaced by the tensor spatial projector
\bea
P^{ijkl}=\left(P^{ik}P^{jl}+P^{il}P^{jk}-P^{ij}P^{kl}\right)/2\,.
\eea

After Fourier transforming, the relevant propagators are 

\be
\left\langle v^jy'_l\right\rangle=\left(\frac{-3}{8\rho}\right)\frac{iP^j_l(\hat k)}{\left[\omega\left(\omega+\frac i{\tau}\right)-\frac15k^2\right]}
\te 

\be
\left\langle \bar x^{jk}y'_{lm}\right\rangle=\left(\frac{-5}{2\tilde\rho}\right)\frac{iP^{jk}_{lm}(\hat k)}{\left[\left(\omega+\frac i{\tau}\right)^2-\frac17k^2\right]}
\te
and the linear fluctuations are

\bea
\left\langle v^jv^k\right\rangle&=&\left(\frac{3\tilde\rho}{40\tau \rho^2}\right)\frac{k^2P^{jk}(\hat k)}{\left[\left(\omega^2-\frac15k^2\right)^2+\frac {\omega^2}{\tau^2}\right]}\\
\left\langle \bar x^{jk}\bar x^{lm}\right\rangle&=&\frac{15}{\tau\tilde\rho}\frac{\left(\omega^2+\frac1{\tau^2}+\frac17k^2\right)P^{jklm}(\hat k)}{\left[\left(\omega^2-\frac{k^2}7-\frac1{\tau^2}\right)^2+\frac{4\omega^2}{\tau^2}\right]}
\tea

At equal times, the fluctuation spectra are

\bea
\left\langle v^jv^k\right\rangle_{t=t'}&=&\left(\frac{3\tilde\rho}{16 \rho^2}\right)\;P^{jk}(\hat k)\\
\left\langle \bar x^{jk}\bar x^{lm}\right\rangle_{t=t'}&=&\left(\frac{15}{2\tilde\rho}\right)\;P^{jklm}(\hat k)
\label{equaltime}
\tea
which do not depend on $\tau$ as expected since in equilibrium for equal times the thermodynamic behaviour is dominant over the dynamical effects.

\subsection{Feynman graphs}
The lowest order contribution to $\Gamma_Q$ is (cfr. \ref{lowest})

\be
\Gamma_Q=\frac{i}2\left(\frac2{5}\tilde\rho \right)^2\int d^4xd^4x'\;\left\langle \left[\left(\dot y'_{jk}-\frac1{\tau}y'_{jk}\right)v^{l} \bar x^{jk}_{,l}\right]\left(x\right) \left[\left(\dot y'_{j'k'}-\frac1{\tau}y'_{j'k'}\right)v^{l'} \bar x^{j'k'}_{,l'}\right]\left(x'\right)\right\rangle_{2PI}
\te
namely
\bea
&&\Gamma_Q=\frac{2i}{25}\tilde\rho^2\int d^4xd^4x'\left\{\left(\frac{\partial}{\partial t}-\frac1{\tau}\right)\frac{\partial}{\partial x'^{l'}}\left\langle y'_{jk}\left(x\right)\bar x^{j'k'}\left(x'\right)\right\rangle\right.  \nn
&&  \left(\frac{\partial}{\partial t'}-\frac1{\tau}\right)\frac{\partial}{\partial x^{l}}\left\langle \bar x^{jk}\left(x\right)y'_{j'k'}\left(x'\right)\right\rangle\left\langle v^{l} \left(x\right)v^{l'} \left(x'\right)\right\rangle\nn
&+&\left(\frac{\partial}{\partial t}-\frac1{\tau}\right)\left(\frac{\partial}{\partial t'}-\frac1{\tau}\right)\left\langle y'_{jk}\left(x\right)y'_{j'k'}\left(x'\right)\right\rangle\nn
&&\left.\frac{\partial^2}{\partial x^{l}\partial x'^{l'}}\left\langle \bar x^{jk}\left(x\right)\bar x^{j'k'}\left(x'\right)\right\rangle\left\langle v^{l} \left(x\right)v^{l'} \left(x'\right)\right\rangle\right\}
\tea

\subsubsection{The self energy}
The self energy

\be
\Sigma^{jk}_{j'k'}=\frac{4i}{25}\tilde\rho^2P^{jk}_{rs}P^{r's'}_{j'k'}\left(\frac{\partial}{\partial t}+\frac1{\tau}\right)\frac{\partial}{\partial x'^{l'}}\frac{\partial}{\partial x^{l}}\left[\left(\frac{\partial}{\partial t'}-\frac1{\tau}\right)\left\langle \bar x^{rs}\left(x\right)y'_{r's'}\left(x'\right)\right\rangle\left\langle v^{l} \left(x\right)v^{l'} \left(x'\right)\right\rangle\right]
\te
In Fourier space
\bea
&&\Sigma^{jk}_{j'k'}\left(k\right)=\left(\frac{3i\tilde\rho^2}{100\tau \rho^2}\right)P^{jk}_{\left(k\right)rs}P^{r's'}_{\left(k\right)j'k'}\left(-ik^0+\frac1{\tau}\right)k_lk_{l'}\nn
&&\int\frac{d\omega d^3p}{\left(2\pi\right)^4}\frac{P^{rs}_{\left(p\right)r's'}\left(\omega+\frac i{\tau}\right)}{\left[\left(\omega+\frac i{\tau}\right)^2-\frac17p^2\right]}\frac{\left(k-p\right)^2P^{ll'}_{\left(k-p\right)}}{\left[\left(\left(\omega-k^0\right)^2-\frac15\left(k-p\right)^2\right)^2+\frac {\left(\omega-k^0\right)^2}{\tau^2}\right]}
\tea
By symmetry $\Sigma^{jk}_{j'k'}=\Sigma P^{jk}_{\left(k\right)j'k'}$, and then

\bea
&&\Sigma=\frac12\Sigma^{ij}_{ij}=\left(\frac{3\tilde\rho^2}{200\tau \rho^2}\right)\left(k^0+\frac i{\tau}\right)\nn
&&\int\frac{d\omega d^3p}{\left(2\pi\right)^4}\frac{P^{r's'}_{\left(k\right)rs}P^{rs}_{\left(p\right)r's'}\left(\omega+\frac i{\tau}\right)}{\left[\left(\omega+\frac i{\tau}\right)^2-\frac17p^2\right]}\frac{\left(k-p\right)^2P^{ll'}_{\left(k-p\right)}k_lk_{l'}}{\left[\left(\left(\omega-k^0\right)^2-\frac15\left(k-p\right)^2\right)^2+\frac {\left(\omega-k^0\right)^2}{\tau^2}\right]}
\tea
We may take $k^i=\delta^i_3k$ and then $p^i=\left(p_{\bot}^a,p^3\right)$, $a=1,2$, whereby

\be
P_{\left(k-p\right)}^{ll'}k_lk_{l'}=\frac{k^2p_{\bot}^2}{\left(k-p\right)^2}
\te
We also find

\be
P^r_{\left(k\right)r'}P^{r'}_{\left(p\right)s}=P^r_{\left(k\right)s}-\frac{p_{\bot}^rp_s}{p^2}
\te

\be
P^{r's'}_{\left(k\right)rs}P^{rs}_{\left(p\right)r's'}=2-2\frac{p_{\bot}^2}{p^2}+\frac14\left(\frac{p_{\bot}^2}{p^2}\right)^2
\te
We can assume approximate isotropy and get

\be
P^{r's'}_{\left(k\right)rs}P^{rs}_{\left(p\right)r's'}\approx\frac45
\te
so now

\be
\Sigma=\left(\frac{3\tilde\rho^2}{250\tau \rho^2}\right)\left(k^0+\frac i{\tau}\right)k^2I
\te 
where

\be 
I=\int\frac{d\omega d^3p}{\left(2\pi\right)^4}\frac{p_{\bot}^2\left(\omega+\frac i{\tau}\right)}{\left[\left(\omega+\frac i{\tau}\right)^2-\frac17p^2\right]\left[\left(\left(\omega-k^0\right)^2-\frac15\left(k-p\right)^2\right)^2+\frac {\left(\omega-k^0\right)^2}{\tau^2}\right]}
\label{integralI}
\te
We give details of the evaluation of the integral (\ref{integralI}) in appendix \ref{Feynman}. From the results there and the previous analysis (\ref{sigmascaling}) we find in the free streaming limit

\be
\Sigma =\frac{2i\tilde\rho}5\left(k^0+\frac i{\tau}\right)^2\left(\frac{\tilde\rho k^3}{\rho^2}\right)\sigma^2\left[\frac{k^0}k\right]
\label{selfen1}
\te 
We shall not need the precise form of the $\sigma$ function in what follows, see Appendix \ref{Feynman}

\subsubsection{The noise kernel}

The noise kernel is

\bea
&&N^{jk}_{j'k'}\left(k\right)=\left(\frac{9\tilde\rho^2}{100\tau^2 \rho^2}\right)P^{jk}_{\left(k\right)rs}P^{j'k'}_{\left(k\right)r's'}\left(\left(k^0\right)^2+\frac1{\tau^2}\right)k_lk_{l'}\nn
&&\int\frac{d\omega d^3p}{\left(2\pi\right)^4}\frac{\left(\omega^2+\frac1{\tau^2}+\frac17p^2\right)P^{rsr's'}_p}{\left[\left(\omega^2-\frac{p^2}7-\frac1{\tau^2}\right)^2+\frac{4\omega^2}{\tau^2}\right]}\frac{\left(k-p\right)^2P^{ll'}_{\left(k-p\right)}}{\left[\left(\left(\omega-k^0\right)^2-\frac15\left(k-p\right)^2\right)^2+\frac {\left(\omega-k^0\right)^2}{\tau^2}\right]}
\tea
We assume $N^{jkj'k'}=NP^{jkj'k'}$. Proceeding as with the self energy, we find

\be
N=\left(\frac{9\tilde\rho^2}{250\tau^2 \rho^2}\right)\left(\left(k^0\right)^2+\frac1{\tau^2}\right)k^2I_N
\te 
where

\be
I_N=\int\frac{d\omega d^3p}{\left(2\pi\right)^4}\frac{\left(\omega^2+\frac1{\tau^2}+\frac17p^2\right)}{\left[\left(\omega^2-\frac{p^2}7-\frac1{\tau^2}\right)^2+\frac{4\omega^2}{\tau^2}\right]}\frac{p_{\bot}^2}{\left[\left(\left(\omega-k^0\right)^2-\frac15\left(k-p\right)^2\right)^2+\frac {\left(\omega-k^0\right)^2}{\tau^2}\right]}
\label{integralN}
\te
In the free streaming limit we find

\be
N=\frac{12\tilde\rho}{5\tau}\left(\left(k^0\right)^2+\frac1{\tau^2}\right)\left(\frac{\tilde\rho k^3}{\rho^2}\right)\mathcal{N}\left[\frac{k^0}k\right]
\label{noisek1}
\te
see Appendix \ref{Feynman} and (\ref{noisescaling}).

\subsection{The spectrum}

The results so far may be summarized by saying that the self energy takes the form (\ref{selfen1}) while the correction to the noise kernel is (\ref{noisek1}). Therefore the corrected symmetric propagator reads

\be 
\langle\bar x^{jk}\bar x_{jk}\rangle=\frac{15}{\tau\tilde\rho}\frac{\left[\left(\left(k^0\right)^2+\frac1{\tau^2}\right)\left[1+\left(\frac{\tilde\rho k^3}{\rho^2}\right)\mathcal{N}\left[\frac{k^0}k\right]\right]+c_T^2k^2\right]}{\left|\left(k^0+\frac i{\tau}\right)^2\left(1+\left(\frac{\tilde\rho k^3}{\rho^2}\right)\sigma^2\left[\frac{k^0}k\right]\right)-c_T^2k^2 \right|^2}
\te
where $c_T^2=1/7$.
We may speculate about the spectrum in a case were loop corrections would be dominant. In that case, we would get
\be 
\langle\bar x^{jk}\bar x_{jk}\rangle=\frac{15\rho^2}{\tau\tilde\rho^2k^3}\frac{\mathcal{N}\left(\frac{k^0}k\right)}{\left(k^{02}+\frac1{\tau^2}\right)\left|\sigma\left(\frac{k^0}k\right) \right|^2}
\te
To compute the equal time correlation we must integrate over $k^0$, which in the free streaming limit adds a further factor of $1/k$, and then at equal times
\be 
\langle\bar x^{jk}\bar x_{jk}\rangle\propto\frac{\rho^2}{\tau\tilde\rho^2k^4}
\label{dressedx}
\te
Remarkably, power law spectra such as this are associated to entropy cascades, with a scale invariant spectrum $k^{-3}$ corresponding to fully developed relativistic turbulence \cite{EC1}.

\section{Final Remarks}

When a nonlinear system is coupled to a random force, mode-mode coupling affects both the inertia of the system and the effective force felt by it. This shows up in such effects as long time tails.

The MSR approach is an efficient tool to incorporate these effects in a consistent way, and takes full advantage of methods developed to treat similar problems in quantum field theory.

In this paper we have demonstrated the MSR approach by applying it to the calculation of the dressed thermal fluctuations of the non hydrodynamic tensor mode of a relativistic viscous fluid.

The existence of such modes is a generic prediction of kinetic theory. Those modes play a leading role in the interaction between fluids and gravitational waves both in cosmological and astrophysical settings.

The dressing by loop corrections changes a flat spectrum for long wavelengths to a power law one at short wavelengths.

We believe these techniques will play an important role in the further analysis of phenomena involving relativistic viscous fluids and electromagnetic and gravitational fields, and look forward to report on further progress soon.

\section{Acknowledgments}

NMG acknowledges financial support by CONICET Grant No. PIP2017/19:11220170100817.

A.K. acknowledges financial support through project uesc 073.11157.2022.0001594-04. 
EC acknowledges financial support from Universidad de Buenos Aires
through Grant No. UBACYT 20020170100129BA,
CONICET Grant No. PIP2017/19:11220170100817CO
and ANPCyT Grant No. PICT 2018: 03684.

A preliminary form of this work was presented at the XL RTFNB - XLII ENFPC 2022, International Institute for Physics, Natal, Rio Grande do Norte, Brazil, September 2022.
\appendix

\section{One loop Feynman graphs}
\label{Feynman}

In this appendix we give further details about the evaluation of the one-loop contributions to the self energy (\ref{integralI}) and the noise kernel (\ref{integralN}).

\subsection{Self energy}
We write eq. (\ref{integralI}) as
\be 
I=\int\frac{d\omega d^3p}{\left(2\pi\right)^4}\;p_{\bot}^2\left(\omega+\frac i{\tau}\right)\left\langle xy\right\rangle_{ret}\left\langle vv\right\rangle_{1}
\label{integralIb}
\te
where

\bea
\left\langle xy\right\rangle_{ret}&=&\frac1{\left[\left(\omega+\frac i{\tau}\right)^2-\frac17p^2\right]}\nn
\left\langle vv\right\rangle_{1}&=&\frac1{\left[\left(\left(\omega-k^0\right)^2-\frac15\left(k-p\right)^2\right)^2+\frac {\left(\omega-k^0\right)^2}{\tau^2}\right]}\nn
\label{vv1}
\tea
$I$ has units of $k$. We factorize

\be
\left\langle vv\right\rangle_{1}=\frac{-i\tau}{2\left(\omega-k^0\right)}\left[\left\langle vv\right\rangle_{ret}-\left\langle vv\right\rangle_{adv}\right]
\label{sigmadec}
\te
where 

\bea
\left\langle vv\right\rangle_{ret}&=&\frac1{\left[\left(\omega-k^0-\frac i{2\tau}\right)^2-\frac15\left(k-p\right)^2+\frac1{4\tau^2}\right]}\nn
\left\langle vv\right\rangle_{adv}&=&\frac1{\left[\left(\omega-k^0+\frac i{2\tau}\right)^2-\frac15\left(k-p\right)^2+\frac1{4\tau^2}\right]}
\tea
Now

\be
\int\frac{d\omega d^3p}{\left(2\pi\right)^4}\;p_{\bot}^2\left(\omega+\frac i{\tau}\right)\frac{-i\tau}{\left(\omega-k^0\right)}\left\langle xy\right\rangle_{ret}\left\langle vv\right\rangle_{adv}=0
\te
so

\be
I=\frac12\int\frac{d\omega d^3p}{\left(2\pi\right)^4}\;p_{\bot}^2\left(\omega+\frac i{\tau}\right)\frac{-i\tau}{\left(\omega-k^0\right)}\left\langle xy\right\rangle_{ret}\left\langle vv\right\rangle_{ret}
\te
It is convenient to write

\be
\frac{\left(\omega+\frac i{\tau}\right)}{\left(\omega-k^0\right)}=1+\left(k^0+\frac {i}{\tau}\right)\frac{\left(\omega+k^0\right)}{\left(\omega^2-k^{02}\right)}
\te
so correspondingly

\be
I=\frac12\left(-i\tau \right)\left[I_1+\left(k^0+\frac {i}{\tau}\right)I_2\right] 
\te
where
\be
I_1=\int\frac{d\omega d^3p}{\left(2\pi\right)^4}\;p_{\bot}^2\left\langle xy\right\rangle_{ret}\left\langle vv\right\rangle_{ret}
\label{integralI1}
\te
\be 
I_2=\int\frac{d\omega d^3p}{\left(2\pi\right)^4}\;p_{\bot}^2\frac{\left(\omega+k^0\right)}{\left(\omega^2-k^{02}\right)}\left\langle xy\right\rangle_{ret}\left\langle vv\right\rangle_{ret}
\label{integralI2}
\te
$I_1$ has units of $k^2$, $I_2$ has units of $k$.

\subsection{Noise kernel}

The noise kernel (\ref{integralN}) may be analyzed in a similar way.  

\be
N=\left(\frac{9\tilde\rho^2}{250\tau^2 \rho^2}\right)\left(\left(k^0\right)^2+\frac1{\tau^2}\right)k^2I_N
\te 
where $I_N$ is dimensionless

\be
I_N=\int\frac{d\omega d^3p}{\left(2\pi\right)^4}\left(\omega^2+\frac1{\tau^2}+\frac17p^2\right)p_{\bot}^2\left\langle xx\right\rangle_1\left\langle vv\right\rangle_1
\te
$\left\langle vv\right\rangle_1$ as in eq. (\ref{vv1}), and

\be
\left\langle xx\right\rangle_1=\frac1{\left[\left(\omega^2-\frac{p^2}7-\frac1{\tau^2}\right)^2+\frac{4\omega^2}{\tau^2}\right]}
\te
$\left\langle vv\right\rangle_1$ may be handled as in eq. (\ref{sigmadec}), and 

\be
\left\langle xx\right\rangle_1=\frac{i\tau}{4\omega}\left\{\left\langle xy\right\rangle_{ret}-\left\langle xy\right\rangle_{adv}\right\}
\te

\bea
\left\langle xy\right\rangle_{ret}&=&\frac 1{\left[\left(\omega+\frac i{\tau}\right)^2-c_T^2p^2\right]}\nn
\left\langle xy\right\rangle_{adv}&=&\frac 1{\left[\left(\omega-\frac i{\tau}\right)^2-c_T^2p^2\right]}
\tea
It follows that

\be
I_N=\frac{\tau^2}4\mathrm{Re}\;\int\frac{d\omega d^3p}{\left(2\pi\right)^4}\frac{\left(\omega^2+\frac1{\tau^2}+\frac17p^2\right)p_{\bot}^2}{\omega\left(\omega-k^0\right)}\left\langle vv\right\rangle_{ret}\left\langle xy\right\rangle_{ret}
\te
The integral is equal to $I_1+I_3$, where $I_1$ is given by (\ref{integralI1}), and $I_3$, which scales as $k^2$,

\be
I_3=\int\frac{d\omega d^3p}{\left(2\pi\right)^4}\frac{\left(\omega k^0+\frac1{\tau^2}+\frac17p^2\right)p_{\bot}^2}{\left(\omega^2-\omega k^0\right)}\left\langle vv\right\rangle_{ret}\left\langle xy\right\rangle_{ret}
\label{integralI3}
\te

\subsection{Computing the integrals}
We now elaborate on the computation of $I_1$. Introducing Feynman parameters \cite{Ramond}

\be
I_1=\int_0^1dx\int\frac{d\omega d^3p}{\left(2\pi\right)^4}\;\frac{p_{\bot}^2}{D_{\Sigma 1}^2}
\label{integralI1b}
\te

\be
D_{\Sigma 1}=\left(\omega+\frac{i}{\tau}\left(1-\frac32x\right)- k^0x\right)^2+\Omega^2+\frac {3ik^0}{\tau}x\left(1-x\right)
\te

\be
\Omega^2=M^2-\left[ xc_V^2+\left( 1-x\right) c_T^2\right] \left( \left( p_z-\delta p_z\right) ^2+p_{\bot}^2\right) 
\te
$c_T=1/\sqrt{7}$, $c_V=1/\sqrt{5}$, 

\be
\delta p_z=\frac{xc_V^2k}{\left[ xc_V^2+\left( 1-x\right) c_T^2\right]}
\te

\be
M^2=x\left(1-x\right)\left[ k^{02}-\frac{c_V^2c_T^2}{\left[ xc_V^2+\left( 1-x\right) c_T^2\right]}k^2\right] -\frac1{\tau^2}2x\left(1-\frac98x\right)
\te
We shift $\omega\to\omega+xk^0$ and $p_z\to p_z+\delta p_z$. 

\be
I_1=\int_0^1dx\int\frac{d\omega d^3p}{\left(2\pi\right)^4}\;\frac{p_{\bot}^2}{\left[\left(\omega+\frac{i}{\tau}\left(1-\frac32x\right)\right)^2+M^2-C^2\left[x\right] p^2 +\frac {3ik^0}{\tau}x\left(1-x\right)\right]^2}
\label{integral1c}
\te
where

\be
C\left[x\right]=\sqrt{xc_V^2+\left( 1-x\right) c_T^2 }
\te
We rescale $p$ and go to polar coordinates

\be
I_1=\frac43\int_0^1\frac{dx}{C^5\left[x\right]}\int\frac{d\omega dp}{\left(2\pi\right)^3}\;\frac{p^4}{\left[\left(\omega+\frac{i}{\tau}\left(1-\frac32x\right)\right)^2+M^2-p^2 +\frac {3ik^0}{\tau}x\left(1-x\right)\right]^2}
\label{integral1e}
\te

Then the integral has (double) poles at

\be
\omega_{\pm}=-\frac{i}{\tau}\left(1-\frac32x\right)\pm i\omega_0
\te

\be
\omega_0=\sqrt{M^2-p^2+\frac {3ik^0}{\tau}x\left(1-x\right)}
\te
If both poles lie on the same half plane, then the integral vanishes.

If $x<2/3$, we close the countour from above, catching the pole at $\omega=\omega_+$

\be
I^<_1=\frac13\int_0^{2/3}\frac{dx}{C^5\left[x\right]}\int\frac{dp\;p^4}{\left(2\pi\right)^2}\;\frac1{\omega_0^3}
\label{integral1f}
\te
The integral is dominated by the value $p_0$ of $p$ such that $\mathrm{Im}\;\omega_+$ is barely above zero. We approximate

\be
\int {dp}\;\frac{p^4}{\omega_0^{\alpha}}=\frac{p_0^3}{\left(\alpha-2\right)\omega_0^{\alpha-2}}
\te
We then have

\be
\int{dp}\;\frac{p^4}{\omega_0^3}\approx \frac{p_0^3}{\omega_0}
\te
Write

\be
\omega_0\left[p_0\right]=\frac1{\tau}\left(1-\frac32x\right)+i\xi
\te
Taking the square of both terms and equating the imaginary parts

\be
\frac{2\xi}{\tau}\left(1-\frac32x\right)=\frac {3k^0}{\tau}x\left(1-x\right)
\te
Observe that $\xi$ is independent of $\tau$

\be
\xi=\frac {3k^0}{2}\frac{x\left(1-x\right)}{\left(1-\frac32x\right)}
\te
and so

\be
p_0^2=M^2+\xi^2-\frac1{\tau^2}\left(1-\frac32x\right)^2
\te
In the $\tau\to\infty$ limit we get

\be
\int\frac{dp\;p^4}{\left(2\pi\right)^2}\;\frac1{\omega_0^3}\approx -i\frac{\left[M^2+\xi^2\right]^{3/2}}{\left(2\pi\right)^2\xi}
\label{LO}
\te
Which is finite when $x\to 0$ but diverges when $x\to 2/3$. This latter divergence is canceled by a divergence with opposite sign coming from the integral with $x>2/3$.

We see that the leading term (\ref{LO}) is imaginary, so the contribution to the noise kernel comes from the next to leading order in the expansion

\be
\frac1{\omega_0}\left[p_0\right]=\frac{-i}{\xi}+\frac1{\tau\xi^2}\left(1-\frac32x\right)+\ldots
\te

$I_2$ (\ref{integralI2}) and $I_3$ (\ref{integralI3}) are computed in the same way. The presence of extra factors in the denominators is handled by adding one more Feynman parameter.

\end{document}